\documentclass[prd,showpacs,showkeys,floatfix,nofootinbib,%%eqsecnum,
               preprint,12pt,tightenlines,fleqn]{revtex4} %for preprint

\usepackage{amsmath,amssymb,revsymb,graphicx,dcolumn}
\newcommand{\beq}{\begin{equation}}
\newcommand{\eeq}{\end{equation}}
\newcommand{\beqa}{\begin{eqnarray}}
\newcommand{\eeqa}{\end{eqnarray}}
\newcommand{\bsubeqs}{\begin{subequations}}
\newcommand{\esubeqs}{\end{subequations}}

\newcommand{\epsilondimless}{e}    %% dimensionless vac energy dens
\begin{document}

%\preprint{KA--TP--19--2011\;(\version)}\vspace*{2mm}
%\preprint{KA--TP--19--2011\;(\today;\;\version)}\vspace*{2mm}
%\preprint{arXiv:xxxxxxxxxxxx,\;KA--TP--19--2011\;(\today;\;\version)}\vspace*{2mm}
%\preprint{arXiv:1107.4063 (\version)}\vspace*{2mm}
%\noindent  arXiv:1107.4063\hfill KA--TP--19--2011\;(\version)\newline\vspace*{2mm}
%
\noindent Phys. Rev. D 85, 023509 (2012) \hfill arXiv:1107.4063\newline\vspace*{2mm}
\title[Inflation and the cosmological constant]
      {Inflation and the cosmological constant\vspace*{5mm}}
\author{F.R. Klinkhamer}
\email{frans.klinkhamer@kit.edu} \affiliation{ \mbox{Institute for
Theoretical Physics, University of Karlsruhe,} Karlsruhe Institute of
Technology, 76128 Karlsruhe, Germany\\}

\begin{abstract}
\vspace*{2.5mm}\noindent
A particular compensation-type solution of the
main cosmological constant problem has been proposed recently,
with two massless vector fields dynamically
canceling an arbitrary cosmological constant $\Lambda$.
The naive expectation is that such a compensation mechanism
does not allow for the existence of an inflationary phase
in the very early Universe.
However, it is shown that certain initial boundary conditions
on the vector fields can in fact
give rise to an inflationary phase.
\end{abstract}

\pacs{04.20.Cv, 98.80.Cq, 98.80.Es}

\keywords{general relativity, inflationary universe, cosmological constant}

\maketitle

\section{Introduction}\label{sec:Intro}

Inflation~\cite{Guth1981,Linde1983}, an epoch of exponential expansion,
may have played an important
role in the evolution of the very early universe
(see Ref.~\cite{precursor-papers}
for an incomplete list of precursor papers and
Ref.~\cite{Mukhanov2005} for further references and discussion).
The mechanism relies, however, on one crucial assumption [stated, for
example, a few lines below Eq.~(3.7) in Ref.~\cite{Guth1981}]:
the minimum of the total scalar potential is set to zero, i.e.,
the corresponding vacuum energy density
(effective cosmological constant $\Lambda$)
is assumed to vanish. In
other words, it is taken for granted that a solution has been found to the
main cosmological constant problem~\cite{Weinberg1988} (CCP1):
why is the present value of
$|\Lambda|^{1/4}$
negligible when compared with the known energy
scales of elementary particle physics? The next cosmological constant
problem (CCP2) is, of course, to explain the measured value
$\Lambda^\text{exp} \sim(2\,\text{meV})^4$,
but this question lies outside the scope of the
present article.

Following an earlier suggestion to consider
vector fields~\cite{Dolgov1985-1997} and using the insights from
the $q$--theory approach~\cite{KV2008-2010} to CCP1,
a special model of two massless vector fields
has been presented in Ref.~\cite{EmelyanovKlinkhamer2011}.
The massless vector fields of the model cancel dynamically
an arbitrary initial (bare) cosmological constant $\Lambda$
without upsetting the local Newtonian dynamics
(a potential problem discussed in Ref.~\cite{RubakovTinyakov1999}).

But, if any cosmological constant can be canceled dynamically,
what happens to inflation in the very early universe?

In order to address this issue,
we investigate the simplest possible
extension of the two-vector-field model by adding a
fundamental scalar field with a quadratic potential,
while keeping an initial cosmological constant $\Lambda$.
The question is, then, whether or not it is possible
to have an inflationary phase of finite duration.

%%\newpage%%tmp
\section{Model}\label{sec:Model}

The model of Ref.~\cite{EmelyanovKlinkhamer2011}
has two massless vector fields $A_\alpha(x)$ and
$B_\alpha(x)$. Now,  a fundamental complex scalar field
$\Sigma(x)$ is added.
Equivalently, it is possible to work with two real scalars
by defining $\Sigma(x)\equiv [\phi_1(x)+i\,\phi_2(x)]/\sqrt{2}$.
The relevant effective action is ($\hbar=c=1$)%
\bsubeqs\label{eq:model}
\beqa\label{eq:model-action}
S_\text{eff}[A,\,B,\,\Sigma,\,g] &=&  - \int{}d^4x\,\sqrt{-g}\;
\Big( \textstyle{\frac{1}{2}}\,(E_\text{Planck})^2\,R
+ \epsilon(Q_A,\,Q_B,\,\Sigma) + \Lambda
\nonumber\\[1mm]
&&
-\partial_{\alpha}\Sigma\;\partial^{\alpha}\Sigma^\star
+U\big(\Sigma\big)
\Big),
\\[2mm]\label{eq:model-QA-QB}
Q_A &\equiv& \sqrt{A_{\alpha;\beta}\;A^{\alpha;\beta}}\,,
\quad%%\\[2mm]
Q_B \equiv \sqrt{B_{\alpha;\beta}\;B^{\alpha;\beta}}\,,
\\[2mm]\label{eq:model-E-Planck}
E_\text{Planck} &\equiv& (8\pi\,G_N)^{-1/2}
                  \approx 2.44\times 10^{18}\:\text{GeV}\,,
\eeqa
for a scalar potential which is real and nonnegative,
$U\big(\Sigma\big)\geq 0$ with $U(|\Sigma_\text{min}|)=0$.
Specifically, the following two functions
$U$ and $\epsilon$ are used:
\beqa
\label{eq:model-scalar-pot}
U\big(\Sigma\big)&=& \,M^2\,|\Sigma|^2 \,,
\\[2mm]
\label{eq:model-espilon}
\epsilon(Q_A,\,Q_B,\,\Sigma) &=& (E_\text{Planck})^4\;
\frac{Q_A^4-Q_B^4}
     {(E_\text{Planck})^8\,\delta_\text{eff}(\Sigma)+Q_A^2\,Q_B^2}\;,
\\[2mm]
\label{eq:model-delta-eff}
\delta_\text{eff}(\Sigma) &\equiv&
\delta\;\frac{|\Sigma|^2}{|\Sigma|^2+(E_\text{Planck})^2\,\eta}\;,
\eeqa
for $0<M\ll E_\text{Planck}$ and (small) positive constants
$\delta$ and $\eta$. The motivation of using the particular
function \eqref{eq:model-delta-eff}
is that, even for a fixed positive value of $\delta$,
the inverse vacuum compressibility $\chi^{-1}$
vanishes if  $\Sigma\to 0$ and the standard local Newtonian dynamics
may be recovered (see Ref.~\cite{EmelyanovKlinkhamer2011}
for further discussion).

The constant $\Lambda$
in the effective action \eqref{eq:model-action} includes the
vacuum-energy-density contributions from the zero-point energies
of the standard-model quantum fields (not shown explicitly).
In principle, this effective cosmological constant $\Lambda$
can be of arbitrary sign and have a magnitude
of order $(E_\text{Planck})^4$.
For further discussion and references on the
effective-action method, see Ref.~\cite[(c)]{KV2008-2010}.

It is also possible to write \eqref{eq:model-action} in terms of
the total potential,
\beqa\label{eq:model-Utot}
U_\text{tot}\big(\Sigma,\,\Lambda\big)&=&
U\big(\Sigma\big)+\Lambda= M^2\,|\Sigma|^2+\Lambda \,.
\eeqa
\esubeqs
As mentioned in the first paragraph of Sec.~\ref{sec:Intro},
this quantity $U_\text{tot}$ has been used in the previous
discussions of inflation, with $\Lambda$ set to zero by hand.
Here, we keep $\Lambda$ arbitrary but introduce vector fields
which have the potentiality to cancel it.
In order to provide this cancellation of the
effective cosmological constant $\Lambda$,
the vector fields must appear in \eqref{eq:model-action} via
the derivative terms contained in $\epsilon$, at least,
within the $q$--theory framework~\cite{KV2008-2010}.

%%\newpage%%tmp
The isotropic \textit{Ansatz}~\cite{Dolgov1985-1997}
for the vector fields $A_{\alpha}(x)$ and $B_{\beta}(x)$,
the scalar $\Sigma(x)$, and the metric $g_{\alpha\beta}(x)$ is%
\bsubeqs\label{eq:Dolgov-Ansatz}
\beqa
A_0 &=& A_0(t)\equiv V(t)\,,\quad\;
A_1=A_2=A_3=0\,,\\[2mm]
B_0 &=& B_0(t)\equiv W(t)\,,\quad
B_1=B_2=B_3=0\,,\\[2mm]
\Sigma &=& \Sigma(t)\,,\\[2mm]
(g_{\alpha\beta})&=&
\text{diag}\big(  1,\,- a(t),\,- a(t),\,- a(t) \big)\,,
\eeqa
\esubeqs
where $t$ is the cosmic time
of a spatially flat Friedmann--Robertson--Walker (FRW) universe,
with cosmic scale factor $a(t)$ and
Hubble parameter $H(t)\equiv [d a(t)/d t]/a(t)$.

Using appropriate powers of the reduced Planck energy
\eqref{eq:model-E-Planck} without additional numerical factors,
the above dimensionful variables can be replaced by
the following dimensionless variables:
\bsubeqs\label{eq:dimensionless-variables}
\beqa
\big\{\Lambda,\, M,\, U,\,\epsilon\,,\,   t,\, H\big\}
&\to&
\big\{\lambda,\,m,\,u,\, \epsilondimless,\,\tau,\,h \big\}\,,\\[2mm]
\big\{Q_A,\, Q_B,\,   V,\, W,\, \Sigma \big\}
&\to&
\big\{q_A,\, q_B,\,   v,\, w,\, \sigma \big\}\,.
\eeqa
\esubeqs
From now on, an overdot stands for differentiation with respect
to $\tau$, for example, $h(\tau)\equiv \dot{a}(\tau)/a(\tau)$.

In terms of these dimensionless variables,
the \textit{Ansatz} \eqref{eq:Dolgov-Ansatz}
reduces the field equations from \eqref{eq:model-action} to
a set of coupled ordinary differential equations
(ODEs) for $v(\tau)$, $w(\tau)$, $h(\tau)$, and $\sigma(\tau)$.
Three of these ODEs have already been given in (3.11)
of Ref.~\cite{EmelyanovKlinkhamer2011}, except that (3.11c)
now contains a dimensionless pressure term from the scalar field,
specifically, $p_\sigma=|\dot{\sigma}|^2-u(\sigma)$.
The fourth ODE is simply the standard FRW Klein--Gordon equation
for $\sigma(\tau)$.

The corresponding Friedmann equation is given by
\bsubeqs
\beqa\label{eq:Friedmann-ODE}
3\, h^2 &=&\lambda+r_\sigma +
\big[\;\widetilde{\epsilondimless}(q_A,\,q_B,\,\sigma)\;
\big]_{q_A=\sqrt{\dot{v}^2+3\,h^2\, v^2},\;\,
q_B=\sqrt{\dot{w}^2+3\,h^2\, w^2}}\,,\\[2mm]
\widetilde{\epsilondimless}
&\equiv&
\epsilondimless -q_A\,\frac{d \epsilondimless}{dq_A}-q_B\,\frac{d \epsilondimless}{dq_B}
=\frac{\big(q_A^2\,q_B^2-3\,\delta_\text{eff}\big)\,\big(q_A^4-q_B^4\big)}
      {\big(\delta_\text{eff}+q_A^2\,q_B^2\big)^2}\;,\\[2mm]
r_\sigma&=&|\dot{\sigma}|^2+u(\sigma)=|\dot{\sigma}|^2+m^2\,|\sigma|^2\,,
\eeqa
\esubeqs
with
$\delta_\text{eff}\equiv\delta\,|\sigma|^2/(|\sigma|^2+\eta)$
from \eqref{eq:model-delta-eff} and $r_\sigma$ corresponding to
the dimensionless energy density from the scalar field.
In conjunction with the four ODEs mentioned
in the previous paragraph, \eqref{eq:Friedmann-ODE} acts as a constraint equation:
if \eqref{eq:Friedmann-ODE}
is satisfied by the initial boundary conditions,
it is always satisfied~\cite{EmelyanovKlinkhamer2011}.

Observe that, in terms of dimensionful variables,
the vacuum energy density $\widetilde{\epsilon}$
on the right-hand side of \eqref{eq:Friedmann-ODE}
differs from the vacuum energy density $\epsilon$
entering the action \eqref{eq:model-action}.
The possible difference of $\widetilde{\epsilon}$
and $\epsilon$ is one of the main results of
$q$--theory (see the original article  Ref.~\cite[(a)]{KV2008-2010}
or the one-page summary of Appendix~A in Ref.~\cite{KV2011-review}).

%%\newpage%%tmp
\section{Results}\label{sec:Results}

Numerical solutions of the reduced field equations
are presented in four figures.
All of these results are obtained from a single model,
specified by the action \eqref{eq:model-action} and
\textit{Ansatz} \eqref{eq:Dolgov-Ansatz},
and are differentiated only by their model parameters
(e.g., $\Lambda$ zero or not) and initial
boundary conditions (e.g., initial vector-field values zero or not).
In principle, the numerical calculation must be performed for
a small but nonzero value of $\eta$
(for example, $\eta=10^{-4}$), but the numerical calculation at
large values of $\tau$ is speeded up by taking the value $\eta=0$.

Figure~\ref{fig:1new} shows an inflationary epoch,
followed by a standard FRW-like expansion phase with
$h \sim (2/3)\,\tau^{-1}$ due to the fact that
$\sigma(\tau)$ rapidly spirals inward
towards the minimum $\sigma_\text{min}=0$.\footnote{Generically,
$\sigma(\tau)$ does not hit $0$ at a finite value of
$\tau$ and the same holds for $\delta_\text{eff}$
from \eqref{eq:model-delta-eff}.
This is the reason for using a complex scalar rather than a
single real scalar which passes through $0$ many times.}
(The main characteristics of this particular type of slow-roll inflation,
for the case of a single real scalar field,
are discussed in Sec.~5.4.1 of Ref.~\cite{Mukhanov2005}.)
Here, the total scalar potential $U_\text{tot}$ in
\eqref{eq:model-Utot}
has its minimum energy fine-tuned to zero, that is $\Lambda=0$.

Figure~\ref{fig:2new} shows that removing the fine-tuning by
changing $\Lambda$
to a positive value leads to eternal inflation without a
subsequent FRW-like phase, due to the presence of a nonzero
value of the vacuum energy density
(effective cosmological constant) even if $\sigma(t)\to 0$.

Figure~\ref{fig:3new} shows that having small but nonzero initial
values of the vector fields leads to the
termination of the inflationary phase by
the eventual vector-field cancellation of the initial
cosmological constant $\Lambda\ne 0$. Changing the value
of $\lambda\equiv \Lambda/(E_\text{Planck})^4$
from $0.01$ to $0.03$ or to $0.0075$ gives similar results.
Returning to $\lambda=0.01$, it has also been verified that setting
$\eta=10^{-4}$ gives essentially the same results
as for $\eta=0$ up to $\tau=450$.

Figure~\ref{fig:4new} shows the absence of a significant
inflationary phase
for large enough initial values of the vector fields,
due to the immediate and complete vector-field cancellation of
$\lambda+u(\sigma)$. Again, it has been verified that setting
$\eta=10^{-4}$ gives essentially the same results
as for $\eta=0$ up to $\tau=10^4$.

Two final comments are as follows.
First, it is seen that $\tau^{-1}\,v(\tau)$ in
Fig.~\ref{fig:3new} or Fig.~\ref{fig:4new}
peaks when $\tau\,h(\tau)$ first drops to $1$,
the position and height of the $\tau^{-1}\,v(\tau)$
peak depending on the initial conditions.
Strictly speaking, this observation also holds for
Fig.~\ref{fig:2new}, with the position of the peak moved off
towards infinity.

Second, extending the numerical solutions of
Figs.~\ref{fig:3new} and \ref{fig:4new} to $\tau=10^6$,
the asymptotic behavior appears to be
$v\sim (q_{A0}/2)\,\tau$, $w\sim (q_{B0}/2)\,\tau$,
and $h \sim 1/\tau$. If confirmed,
this asymptotic behavior would correspond to a different branch
than the one found numerically in Ref.~\cite{EmelyanovKlinkhamer2011}
(for the theory without scalars)
and would in fact correspond to the standard
$q$--theory branch~\cite[(c)]{KV2008-2010} with constant vacuum
variables $q_A\equiv \sqrt{\dot{v}^2+3\,h^2\, v^2}=q_{A0}$
and       $q_B\equiv \sqrt{\dot{w}^2+3\,h^2\, w^2}=q_{B0}$.

%%\newpage%%tmp
\section{Discussion}\label{sec:Discussion}

\begin{figure*}[th]%%[p][t]  %%figX_v211.eps-->figX_v3.eps-->figX_v4.eps
%\vspace*{-1.5cm}
\begin{center}
\hspace*{-6mm}
\includegraphics[width=1.05\textwidth]{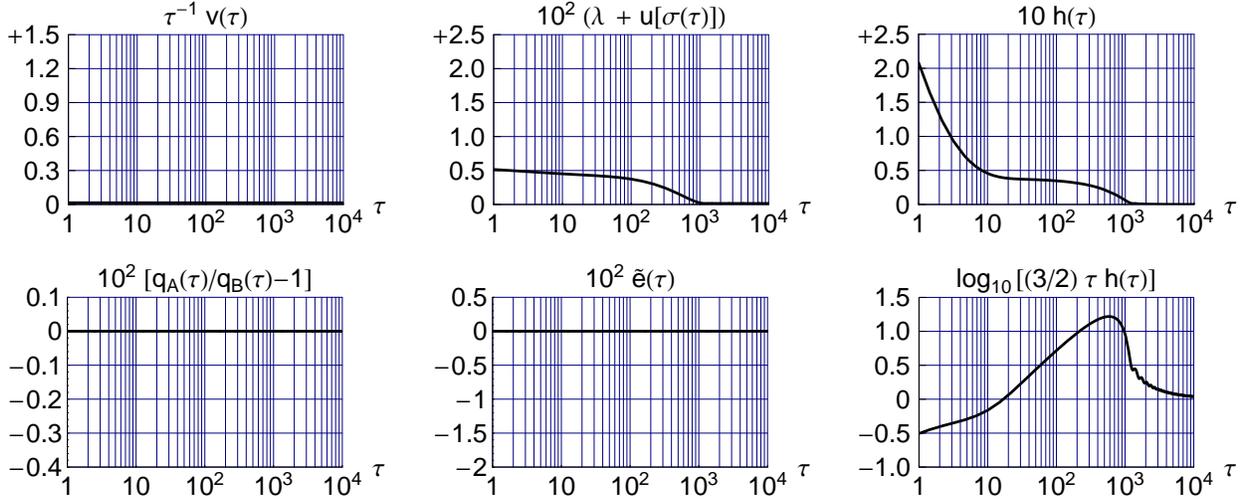}
\end{center}
\vspace*{-5mm}
\caption{Numerical solution of the reduced field
equations for model \eqref{eq:model} and
\textit{Ansatz} \eqref{eq:Dolgov-Ansatz}.
The dimensionless model parameters are $m=0.01$, $\delta=10^{-6}$,
$\eta=0$, and $\lambda=0$. The boundary conditions are $a(1)=1$,
$\{\varphi_1(1),\,\dot{\varphi}_1(1),\,\varphi_2(1),\,\dot{\varphi}_2(1)\}$
$=$ $\{10,\, -0.25,\, 0,\, -0.433013 \}$,
and $v(1)=\dot{v}(1)= w(1)=\dot{w}(1)=0$.
[The real scalars $\varphi_n$ are defined by
$\sigma\equiv (\varphi_1+i\,\varphi_2)/\sqrt{2}$.]
The value of $h(1)$ follows from the Friedmann equation
\eqref{eq:Friedmann-ODE}. With these boundary conditions,
the reduced field equations have the
exact solution $v(\tau)=w(\tau)=0$ for $\tau \geq 1$.
The top right panel shows an $h(\tau)$ plateau corresponding to an
inflationary phase.
The bottom right panel shows that, long after inflation,
$h(\tau)$ asymptotically goes as $(2/3)\,\tau^{-1}$.
\vspace*{0cm} } \label{fig:1new}
\end{figure*}

\begin{figure*}[h]%%[p][t]
\vspace*{0cm}
\begin{center}
\hspace*{-6mm}
\includegraphics[width=1.05\textwidth]{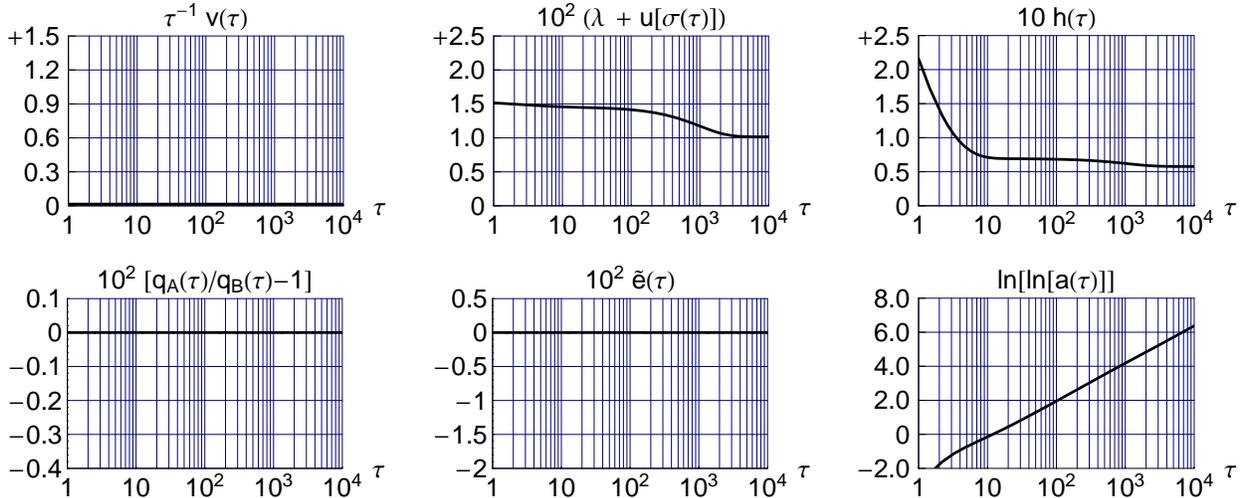}
\end{center}
\vspace*{-5mm}
\caption{Same as Fig.~\ref{fig:1new}, but now with nonzero
cosmological constant, $\lambda=0.01$.
Same boundary conditions as Fig.~\ref{fig:1new},
so that the exact solution $v(\tau)=w(\tau)=0$ persists.
The Friedmann equation \eqref{eq:Friedmann-ODE} gives
the asymptotic value $h(\infty)=\sqrt{\lambda/3}\approx 0.05774$.
\vspace*{-4cm} } \label{fig:2new}
\end{figure*}

\begin{figure*}[th]%%[p][t]
%\vspace*{-1.5cm}
\begin{center}
\hspace*{-6mm}
\includegraphics[width=1.05\textwidth]{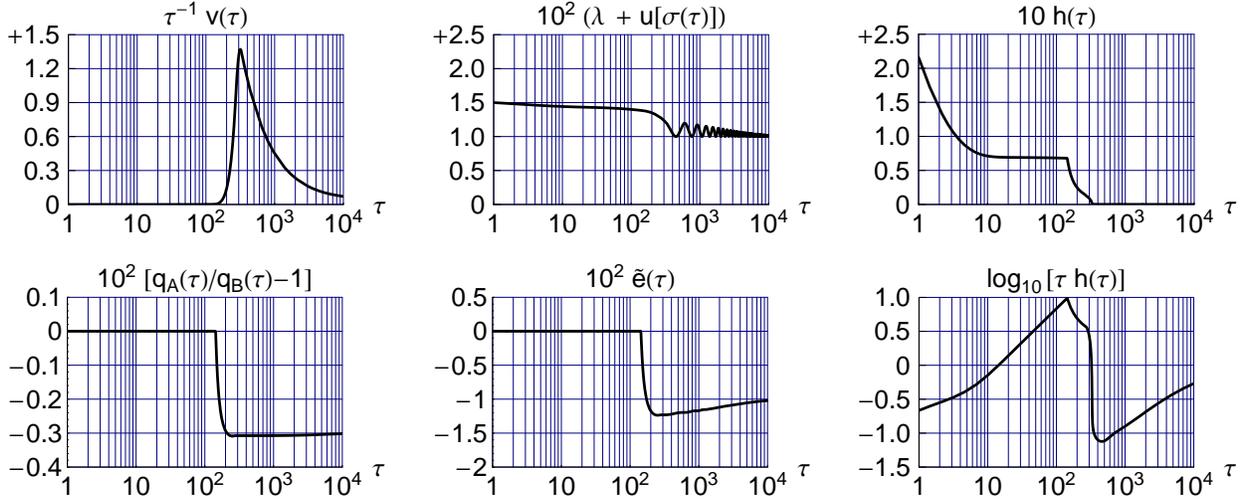}
\end{center}
\vspace*{-5mm}
\caption{Same parameters and boundary conditions as Fig.~\ref{fig:2new}
(e.g., $\lambda=0.01$), except for small but nonzero starting values of
$v$ and $w$. Specifically, the
boundary conditions at $\tau=1$ are
$\{a,\,v,\,\dot{v},\,w,\,\dot{w},\,h,\,
  \varphi_1,\,\dot{\varphi}_1,\,\varphi_2,\,\dot{\varphi}_2\}$
$=$
$\{1,\,2\times 10^{-4},\,2\times 10^{-4},\,2\times 10^{-4},\,2\times 10^{-4}$, $0.216025$,
$10,\, -0.25,\, 0,\, -0.433013 \}$.
%
%
%{v,vdot,w,wdot,h,phi1,phi1dot,phi2,phi2dot}[tmin]                %%v210
%{0.0002, 0.0002, 0.0002, 0.0002, 0.216025, 10., -0.25, 0., -0.433013}
%
The total vacuum energy density entering the right-hand side
of the Friedmann equation \eqref{eq:Friedmann-ODE} is given by the
sum of the two panels in the middle column and vanishes
asymptotically.
\vspace*{1\baselineskip}} \label{fig:3new}
\end{figure*}

\begin{figure*}[h]%%[p][t]
\vspace*{0cm}
\begin{center}
\hspace*{-6mm}
\includegraphics[width=1.05\textwidth]{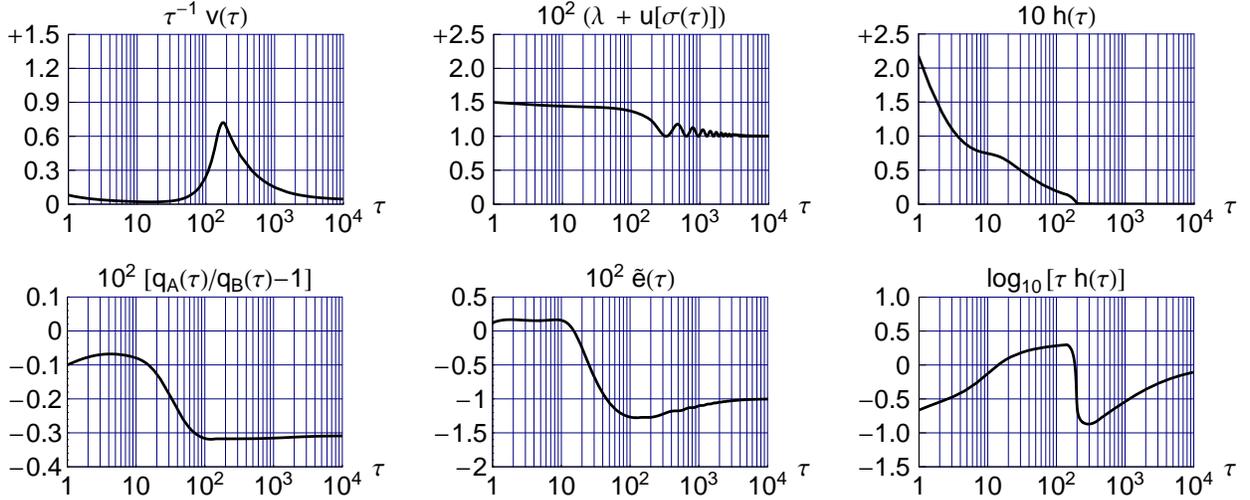}
\end{center}
\vspace*{-5mm}
\caption{Same parameters and boundary conditions as Fig.~\ref{fig:3new}
(e.g., $\lambda=0.01$), except for relatively large starting values
of $v$ and $w$. Specifically, the boundary conditions at $\tau=1$ are
$\{a,\,v,\,\dot{v},\,w,\,\dot{w},\,h,\,
   \varphi_1,\,\dot{\varphi}_1,\,\varphi_2,\,\dot{\varphi}_2\}$
$=$
$\{1,\,0.0799201,\, 0.01998,\, 0.08,\, 0.02$, $0.216961$,
$10,\, -0.25,\, 0,\, -0.433013 \}$.
%
%{v,vdot,w,wdot,h,phi1,phi1dot,phi2,phi2dot}[tmin]   %%v150
%{0.0799201, 0.01998, 0.08, 0.02, 0.216961, 10., -0.25, 0., -0.433013}
%
\vspace*{-4cm} }
\label{fig:4new}
\end{figure*}

The results of Fig.~\ref{fig:4new} were to be expected.
The surprising (and encouraging?) results are
those of Fig.~\ref{fig:3new}, with an inflationary phase of
some 10 $e$-foldings of $a(\tau)$ for the model and
parameters chosen. (Different initial conditions have been
seen to give some 30 $e$-foldings and Fig.~\ref{fig:2new}
can be interpreted as having infinitely many $e$-foldings.)
Qualitatively, the Hubble parameter $h(\tau)$ of the top right panel
in Fig.~\ref{fig:3new} ($\lambda \ne 0$)
resembles that of Fig.~\ref{fig:1new} ($\lambda = 0$),
even though the detailed behavior differs as shown by
the respective  bottom right panels.

The results of Fig.~\ref{fig:3new} are, of course, only exploratory.
It remains, for example, to analyze the nonlinear dynamics
displayed in Fig.~\ref{fig:3new}
and to rigorously establish the $\tau\to\infty$ limit for the
$\eta=10^{-4}$ case, corresponding respectively
to an early phase with inflation\footnote{Possible
observable effects may come from the dynamics of density perturbations
with near-horizon wavelengths (cf. Ref.~\cite{Mukhanov2005}),
as the dynamics can be expected to be modified
by the interaction between the scalar (inflaton) field
and the vector fields needed for the cancellation
of the \mbox{bare cosmological constant $\Lambda$.}}
and a late (FRW-like) phase with
standard local Newtonian dynamics.

The main result of this article is that, in principle,
it appears to be possible to have both an early phase with inflation
and a late phase with a dynamically canceled cosmological constant $\Lambda$.
The details of the model are of secondary importance.
What matters is the general mechanism which relies on the dynamics
of the massless vector fields (or possibly massless tensor fields).
The physical origin of these massless vector (tensor) fields
needs to be clarified.

\newpage
%\vfill

\newpage
%\vfill

\newpage%%tmp
\section*{\hspace*{-4.5mm}ACKNOWLEDGMENTS}
\noindent
It is a pleasure to thank G.E. Volovik for numerous
discussions on the cosmological constant problem,
V. Emelyanov for a valuable suggestion, and
S. Thambyahpillai for helpful comments on the manuscript.

\end{document}